\documentclass[journal=jacsat,manuscript=article]{achemso}

\usepackage[utf8]{inputenc}
\usepackage[T1]{fontenc}
\usepackage{graphicx}
\usepackage{amsmath,amssymb}
\usepackage{physics}
\usepackage{color}
\usepackage{multirow}
\usepackage{float}
\usepackage{braket}
\usepackage{adjustbox}
\usepackage{booktabs, tabularx}
\newcolumntype{C}{>{\centering\arraybackslash}X}
\usepackage[normalem]{ulem}
\usepackage{placeins}

\newcommand*{\vt}[1]{\textcolor{black}{ #1}}
\newcommand*{\vtnew}[1]{\textcolor{black}{ #1}}
\newcommand*{\patra}[1]{\textcolor{black}{ #1}}

\author{Sanjoy Patra}
\affiliation{Solid State and Structural Chemistry Unit, Indian Institute of Science, Bangalore, Karnataka 560012, India}
\author{Atandrita Bhattacharyya}
\affiliation{Solid State and Structural Chemistry Unit, Indian Institute of Science, Bangalore, Karnataka 560012, India}
\author{Ch. Mudasar Hussain}
\affiliation{School of Physical Sciences, Jawaharlal Nehru University, Delhi, New Delhi 110067, India}
\author{Vijay P. Singh}
\affiliation{School of Physical Sciences, Jawaharlal Nehru University, Delhi, New Delhi 110067, India}
\author{Supriyo Santra}
\affiliation{School of Chemical Sciences, Indian Association for the Cultivation of Science, Kolkata 700032, India}
\author{Debashree Ghosh}
\affiliation{School of Chemical Sciences, Indian Association for the Cultivation of Science, Kolkata 700032, India}
\author{Pritam Mukhopadhyay}
\affiliation{School of Physical Sciences, Jawaharlal Nehru University, Delhi, New Delhi 110067, India}

\author{Vivek Tiwari}
\email{vivektiwari@iisc.ac.in}
\affiliation{Solid State and Structural Chemistry Unit, Indian Institute of Science, Bangalore, Karnataka 560012, India}

\title{Intramolecular Singlet Fission Through a Coherently Coupled Excimer-like Intermediate}

\begin{document}
	
\begin{abstract}
Singlet Fission (SF) into two triplets offers exciting avenues for high-efficiency photovoltaics and optically initializable qubits. While the chemical space of SF chromophores is ever-expanding, the mechanistic details of electronic-nuclear motions that dictate the photophysics are unclear. Rigid SF dimers with well-defined orientations are necessary to decipher such details. Here, using polarization-controlled white-light two-dimensional and pump-probe spectroscopies, we investigate a new class of contorted naphthalenediimide dimers, recently reported to have a favorable intramolecular SF (iSF) pathway. 2D cross-peaks directly identify the two Davydov components of the dimer along with strongly wavelength-dependent $TT_1$ formation kinetics depending on which Davydov component is excited, implicating a coherently coupled intermediate that mediates iSF. Enhanced quantum beats in the $TT_1$ photoproduct suggest that inter-chromophore twisting and ruffling motions drive the $\sim$200 fs evolution towards an excimer-like intermediate and its subsequent $\sim$2 ps relaxation to the $TT_1$ photoproduct. Polarization anisotropy directly tracks electronic motion during these steps and reveals surprisingly minimal electronic reorientation with significant singlet-triplet mixing throughout the nuclear evolution away from the Franck-Condon geometry towards relaxed $TT_1$. \vtnew{The observations of coherent excimer-like intermediate and significant singlet-triplet mixing throughout the iSF process need to be carefully accounted for in the synthetic design and electronic structure models for iSF dimers aiming for long-lived high-spin correlated triplets.}
\end{abstract}

\maketitle


\section{INTRODUCTION}
Design of new chromophores to understand the interplay of electronic states with nuclear and spin evolution that causes singlet fission\cite{Smith2010Singlet} (SF) of the excited singlet into two free triplets \patra{($T_1$)} has garnered significant interest due to possible applications in solar light harvesting, and more recently as molecular qubits\cite{Dill2023Entangled,Smyser2020Singlet} and enhancement of nuclear magnetic resonance (NMR) sensing through electronic spin polarization\cite{Kawashima2023Singlet}. Acene thin films are the archetypal examples of solution-processed intermolecular SF materials where experimental reports and theoretical proposals on SF mechanism have ranged from direct or coherent generation\cite{Zhu2012,Zhu2017,Turner2017,Musser2024} of the correlated triplet pair $TT_{1}$ from the locally excited ($LE$) singlet, to that mediated\cite{Phys2014Microscopic,Margulies2017Direct,Hong2020Efficient,Miller2017Modeling} by higher-lying charge-transfer ($CT$) states. Electronic couplings and adiabatic evolution\cite{Feng2013Fission} of the wavefunction is often insufficient to explain the coherent photoexcitation of the $TT_{1}$ state, thus leading\cite{Morrison2017Evidence,Tempelaar2018Vibronic,Bhattacharyya2023Low} to the proposed role of vibronic couplings in mediating this process. \\

Structurally rigid intramolecular SF (iSF) dimers are a particularly interesting test bed for understanding the mechanistic aspects of SF because exquisite control over electronic coupling matrix elements becomes possible by synthetically tuning molecular symmetry\cite{Gilligan2019Using}. One of the pitfalls of structurally well-defined dimers is that excessive orbital overlaps due to center-to-center distances of only 3--4 \AA{} can also lead to the undesired formation of excimers\cite{Liu2015Synthesis} and low-energy $CT$ traps\cite{Margulies2016Enabling}. Consequently, the interplay of molecular geometry, electronic couplings and possibly even nuclear motions that differentiates one photophysical outcome from another becomes quite challenging to decipher. For example, $CT$ states are reported to both, mediate\cite{Margulies2017Direct,Hong2020Efficient} and sometimes inhibit\cite{Margulies2016Enabling} SF. In the context of iSF, excimer formation has been \vtnew{generally reported as an undesired pathway\cite{Hong2020Efficient,Liu2015Synthesis,Jadhav2022Modulating,KimWurthner2022} that is fast enough to effectively compete against the slower $TT_1$ formation kinetics. Excimer formation is reported to be enabled by structural relaxation\cite{Hong2020Efficient,Feng2016On,Kim2021,Kim2022} on the excited state leading to a frustrated photodimerization which increases\cite{Brown2014Direct}  with electronic coupling. Only fairly recent works report\cite{Guldi2019,KimWurthner2022} excimer-mediated iSF but with slow $TT_1$ formation on $\sim$100 ps and slower timescales.  When SF is intermolecular, excimer-mediated SF is reported to be similarly slower -- 2.2 ps excimer formation and 22 ps $TT_{1}$ formation in diketopyrrolopyrrole thin films\cite{Mauck2016Singlet}, 46 ps $TT_{1}$ formation in $\pi$-stacked terrylenediimide (TDI) thin films\cite{Margulies2017Direct}, and 2--7 ps $TT_{1}$ formation in TIPS-Tc nanoaggregates\cite{Thampi2018Elucidation} where in the latter two cases the excimer formation was faster than the few hundred femtosecond instrument response. Kim, W\"{u}rthner and co-workers have elucidated\cite{Kim2021, Kim2022,KimWurthner2022} the structural dynamics associated with the photophysics of excimer formation. The corresponding electronic dynamics carries similar mechanistic significance and is one of the focus of this work.} \\

Theoretical models from Krylov\cite{Feng2016On} and co-workers have accounted for excimer formation to argue that structural relaxation of the photoexcited singlets in flexible staggered tetracene dimers towards an excimer minima can in fact make the $TT_{1}$ adiabat energetically accessible compared to the Franck-Condon (FC) geometry, but often at the expense of inhibiting the $TT_{1}$ formation by enhancing non-radiative relaxation of the excimer. The latter parasitic pathway is in line with the observations\cite{Hong2020Efficient} of Hong et al. who report that a structurally relaxed iSF dimer (excimer, $EX$) can borrow $S_{1}$ oscillator strength and relax directly to the ground state. Schatz and co-workers have extended the $CT$-mediated picture of SF to propose\cite{Miller2017Modeling} a state vector picture for various configurations that may be present in iSF dimers such that the excimer $EX$ is defined as having a borrowed electronic character from $LE$ and $CT$ states and possibly also from the $TT_{1}$ state depending on their relative adiabatic energies. At the same time, recent observations of near-instantaneous generation of coherently mixed $LE-CT$ intermediates\cite{Lin2022Accelerating} in perylenediimide (PDI) trimers, vibronically coherent formation of $LE-TT$ intermediates\cite{Stern2017Vibronically} during intermolecular SF in tetracene thin films, and $EX-CT$ non-adiabatic intermediates\cite{Hong2022Ultrafast} during symmetry breaking charge separation (SBCS) dimers prompted us to explore the following questions in the context of iSF -- 1. Can coherent intermediates, formed as a result of strong electronic/vibronic couplings, overcome the unavoidable excimer traps to form $TT_{1}$ in strongly coupled iSF dimers? 2. \vt{Strongly coupled sites and wavelength dependent iSF} is theoretically expected\cite{Bhattacharyya2023Low} in rigid iSF dimers but both features have not been directly confirmed in any iSF system till date. 3. What is the electronic reorientation involved during the evolution of a photoexcited singlet towards these intermediates? Given the near-instantaneous generation of above intermediates, this question remains unclear and has not yet been probed directly. \\

In relation to the latter question, our recent work\cite{Bhattacharyya2023Low} on a vibronic dimer picture of iSF with explicit role of low and high-frequency vibrations reveals that given the large nuclear reorganization energies on the $TT_{1}$ state, only a small amount of electronic mixing between $LE$ and $TT_{1}$ -- direct or mediated -- is sufficient to lead to vibronically enhanced $LE-TT_{1}$ mixing. Given this expectation, rich dynamical effects such as coherent generation of a $LE-TT_{1}$ intermediate and wavelength dependent iSF rates depending on the excitation of lower or upper Davydov component of the dimer, are to be expected but not reported for any iSF system till date. One of us\cite{Bansal2022highly} recently reported a new class of naphthalenediimide (NDI) dimers where a contortion of the NDI skeleton along with a relative rotation of 20$^\circ$ and 24$^\circ$ along the long and short axes, respectively, avoids excimer formation even with center-to-center distance of 3.3 \AA  between the rings, such that iSF becomes a favorable pathway. Sub-picosecond iSF was confirmed by \vtnew{non-impulsive} pump-probe (PP) experiments and through a Ruthenium-based triplet sensitizer with NDI dimer triplets directly probed by microsecond PP experiments. Absence of any $CT$ intermediates was confirmed by absence of radical anion and cation NDI bands\cite{Ajayakumar2012Core} in the PP spectra. The questions in the preceding paragraph are best explored in rigid strongly coupled iSF dimers and prompted us for a deeper mechanistic investigation of iSF in this system. \\

Here, using a combination of white-light, polarization-controlled two-dimensional electronic spectroscopy (2DES) and impulsive PP experiments we show that iSF in NDI dimers proceeds through a coherently coupled intermediate that forms on a $\sim$200 fs timescale. This coherent picture of iSF is evidenced by our measurements of 2DES cross-peaks (CPs) and the reduced electronic polarization anisotropy, where both observations are consistent with strong electronic correlations between the chromophore sites. \vtnew{Interestingly, the formation and rate of this $[S_1 + TT_1]$ intermediate is not excitation wavelength or solvent polarity dependent.} However, its subsequent relaxation to the $TT_{1}$ state is strongly wavelength dependent -- 2 ps when upper Davydov component is excited versus 0.6 ps when the lower Davydov component is excited. Enhanced quantum beats observed in the $TT_{1}$ photoproduct suggest that inter-chromophore twisting and ruffling motions likely drive the excited state evolution towards an excimer-like intermediate and subsequent to a relaxed $TT_{1}$ state. By directly tracking the electronic reorientation during $S_{1}$, $[S_1 + TT_1]$ intermediate and $TT_{1}$ formation, we find that there is surprisingly little electronic reorientation during these steps and that strong electronic correlations between the chromophore sites and significant singlet-triplet mixing are maintained throughout the $TT_1$ formation process. \vtnew{To best of our knowledge, the observations of 2DES cross-peaks, coherent excimer-like intermediate, and strongly wavelength-dependent iSF with surprisingly minimal electronic reorientation throughout the process have not been reported before in any iSF system and elucidate the deeper mechanistic landscape that underpins singlet fission. Our observations also demonstrate that tracking electronic transition dipole reorientation during $TT_{1}$ formation can provide powerful insights to advance the synthetic design and electronic structure models of strongly coupled iSF dimers aiming to synthetically tune long-lived high-spin triplet states. }

\section{RESULTS AND DISCUSSION}

\subsection{Linear Measurements}
\begin{figure*}[!ht]
	\centering
	\includegraphics[width=5 in]{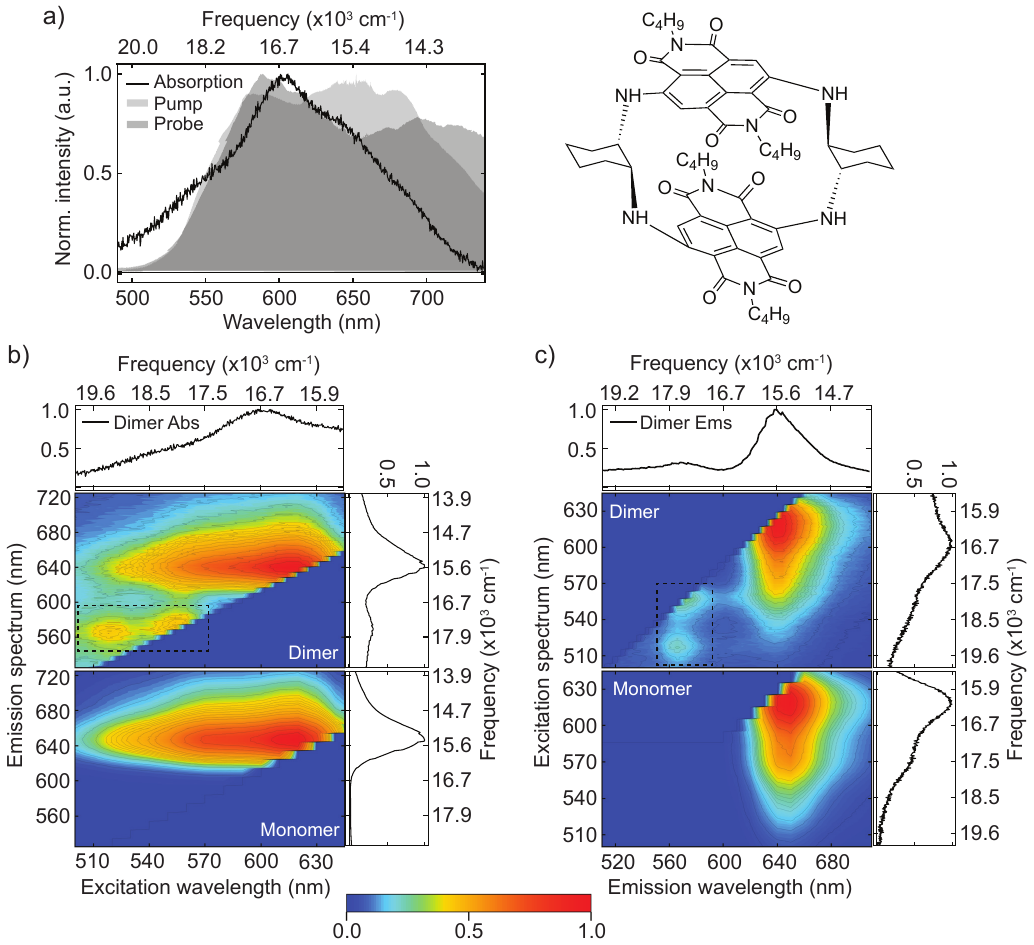}
	\caption{ \footnotesize \textbf{Linear excitation-emission maps reveal low-oscillator strength emissive states.} (\textbf{a}) Linear absorption spectrum of NDI dimer in a 1:1(v/v) mixture of solvents acetonitrile and dichloromethane (ACN and DCM, respectively). The spectra are overlaid with the pump and probe white-light laser spectrum used for the impulsive PP and 2DES experiments. The NDI dimer molecule is shown on the right. (\textbf{b}) Emission spectrum collected as a function of excitation wavelength for dimer (top) and monomer (bottom). The 1D spectrum on the top inset shows the absorption spectrum of the dimer and the side panel shows the respective integrated emission spectrum. (\textbf{c}) Excitation spectrum as a function of emission collection wavelength. The top panel is the integrated emission spectrum of the dimer while the side panel is the respective absorption spectrum. 
	}
	\label{fig:fig1}
\end{figure*}

The linear absorption spectrum of the NDI dimer is overlaid with the pump and probe spectrum in \vt{Figure~\ref{fig:fig1}a}. Unlike the conformationally heterogeneous\cite{Zirzlmeier2015Singlet, Guldi2019} and significantly weakly coupled\cite{Margulies2016Enabling} iSF dimers \vt{which are dominated by FC-like progressions even in the dimer, distinct peaks and shoulders are seen at 600 nm and 650 nm, respectively, which are conclusively confirmed to be the Davydov components -- upper and lower excitons of the dimer -- through 2DES measurements} (\textit{vide infra}). The shoulder at 550 nm is likely a FC shoulder resulting from the overlapping progressions of the two Davydov components. The single crystal X-ray crystallographic structure of the dimer that we reported earlier showed a center-to-center distance of 3.3~\AA\ which was of the order of the 3.4~\AA\ van der Waals radii of the C atoms, suggesting strong orbital overlaps and possibly a significant role for $CT$ states. To investigate this further, we first performed linear emission measurements as a function of excitation wavelength as shown in \vt{Figure~\ref{fig:fig1}b}. In order to rule out any fluorescence reabsorption artifacts, the sample concentration for the emission measurements was diluted to 700 nM such that the average molecular separation for molecular number density of 4.2$\times$10$^{20}$ molecules/cm$^{3}$ was $\sim$21.4$\times$ larger than the \patra{F\"{o}rster} critical radius\cite{Forster195910th,Cho2013Absolute,Tiwari2018Strongly} for energy transfer. The emission spectrum at 700 nM versus a higher concentration of 940 $\mu$M is overlaid in \vt{Figure S2b} and shows no reabsorption effects. Compared to the monomer emission spectrum (\vt{Figure~\ref{fig:fig1}b}, bottom panel), excitations at around 525 nm and 555 nm result in emission peaks at 567 nm that are 4$\times$ weaker than the main emission band at 640 nm. This pattern is further confirmed in the complementary measurement of excitation spectrum as a function of emission collection wavelength (\vt{Figure~\ref{fig:fig1}c}). The weak peaks at 520 nm and 555 nm are seen here as well for collection wavelengths of 568 nm. These observations suggest that the blue part of the absorption spectrum (\vt{Figure~\ref{fig:fig1}a}) below 550 nm consists of weakly emissive states with well-defined emission profiles unlike the low-oscillator strength broad emission known\cite{Young2020Mixed, SpanoExcimer2022} in excimers. Given the strong electronic couplings expected in this system, these are likely to be $CT$ states with borrowed bright character from $LE$ states. Significant intensity borrowing between $LE-CT$ states was indeed predicted by previous electronic structure calculations\cite{Bansal2022highly} on this system. Additionally, TD-DFT calculations\cite{Conrad2019Controlling} from Zhu and co-workers on a contorted PDI dimer show that the singlet oscillator strength is significantly distributed among higher-lying charge-transfer $CT$ and singlet type states (see Section S8 of ref.\cite{Conrad2019Controlling}). This also explains the reduced molar extinction coefficient\cite{Bansal2022highly} seen in the contorted NDI dimer (5530 M$^{-1}$cm$^{-1}$) as compared to the monomer (27440 M$^{-1}$cm$^{-1}$).

\subsection{2DES reveals Davydov splitting and excitation-wavelength dependent $TT_1$ formation}
\begin{figure*}[!ht]
	\centering
	\includegraphics[width=5 in]{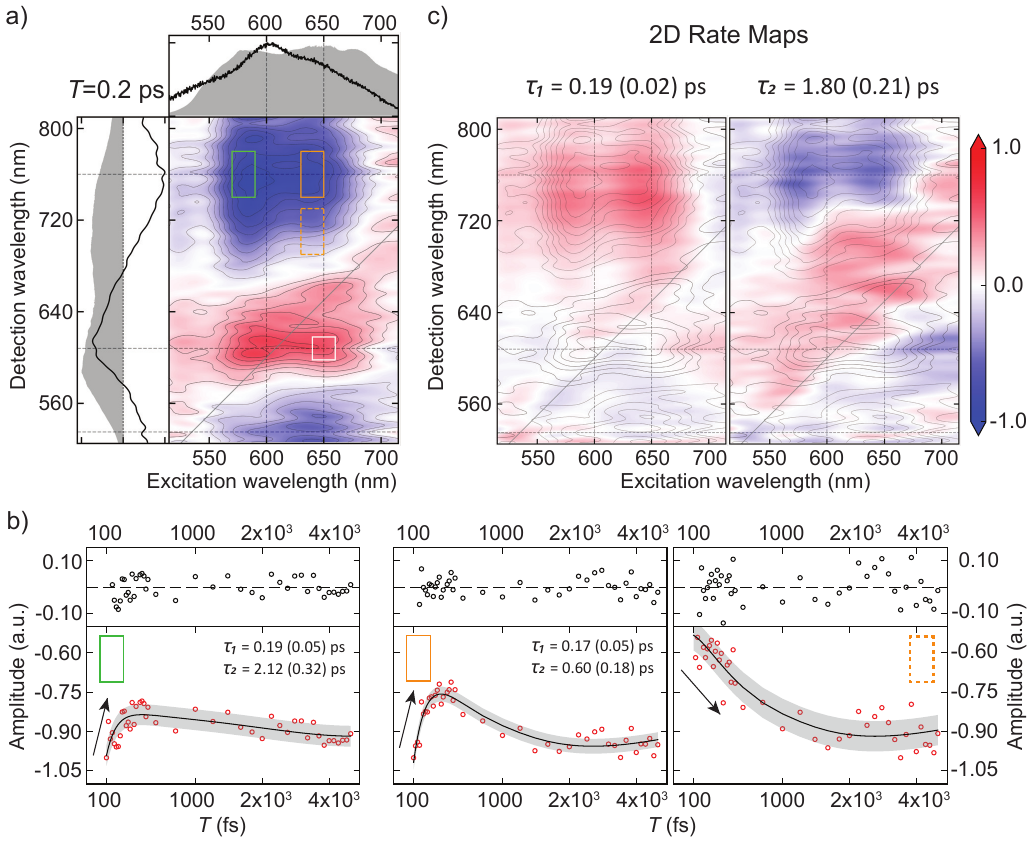}
	\caption{ \footnotesize \textbf{2DES reveals Davydov splitting and excitation wavelength dependent $TT_1$ formation.} (\textbf{a}) 2D spectrum of the NDI dimer at $T$ = 0.2 ps. The top panel shows the absorption spectrum (black) overlaid with the pump spectrum (gray shaded). The side panel shows the spectrally integrated spectrum versus the detection wavelength, $\lambda_t$, overlaid with the probe spectrum (gray shaded). Contours are drawn at 10\%  to 100\% in 10\% intervals. The positive ground state bleach (GSB) signal is shown in red while the negative excited state absorption (ESA) signal is shown in blue. (\textbf{b}) (left) Dynamics in the ESA region denoted by the \patra{green} rectangle in panel A. (middle) Dynamics in the ESA region denoted by the \patra{yellow} rectangle in panel A. (right) Dynamics in the ESA region denoted by dashed \patra{yellow} rectangle in panel A. For panel b (left and middle), the data was collected until 5 ps and the spectrally integrated 2D signals are fitted to a freely floated 3-exponential model where the first two time-constants are shown in the figure. For panel b (right), the time constants of the fit function for the middle panel are kept fixed and only the amplitudes are floated. The accompanying top panel shows the respective residuals. \patra{The $\pm {\sigma}$ error band on the fit is calculated by averaging N=6 trials of \vt{$T$} = 0.5 ps 2D spectrum where collection of individual trials was interleaved over the duration of the full scan.} \patra{The arrows denote the 200 fs decay or rise, that is followed by the slower rise of the ESA signal.} (\textbf{c}) 2D rate maps obtained by global fitting of all 2D pixels to a 3-exponential model. The obtained time constants are shown above the respective maps. The rate maps are normalized by the sum of the maxima of two rate maps. Global rate analysis neglects the excitation wavelength dependence seen in panel b (left versus middle). Contours are drawn from 10\% to 100\% in 10\% intervals.
	}
	\label{fig:fig2}
\end{figure*}

A 2DES map correlates excitation and detection frequencies of a system in the form of a 2D contour map of detection versus excitation axis which evolves as a function of pump-probe waiting time $T$. Our 2DES experiments are conducted in a partially collinear pump-probe geometry where the pump-pulse pair is generated through a common path interferometer. The collected signal measures the absorptive changes in the transmitted probe spectrum. Further experimental details are provided in \vt{Section S1}. To investigate the nature of the absorption peaks at 600 nm and 650 nm (\vt{Figure~\ref{fig:fig1}a}), we conducted 2DES measurements of kinetic rate maps (RMs). \vt{Figure~\ref{fig:fig2}a} shows the NDI dimer spectrum at $T$ = 0.2 ps. A clear cross-peak (CP) with excitation wavelength 650 nm and detection wavelength 608 nm, that is, between the main absorption peak and its red shoulder is seen. This is denoted by the white square in \vt{Figure~\ref{fig:fig2}a}. It is important to note that this is a positive upper 2D cross-peak ($\lambda_t < \lambda_\tau$, CP$_U$) which arises only for strongly correlated electronic transitions where bleaching one transition also affects the other, that is, a shared ground state between electronic states. Thus, the presence of a positive CP$_U$ is a direct evidence for the main peak and the red shoulder in the absorption spectrum being Davydov components with a \vt{$\sim$1040 cm$^{-1}$} energetic splitting suggesting large electronic coupling between the sites. Note that unlike conformationally heterogeneous\cite{Liu2015Synthesis,Zirzlmeier2015Singlet,Guldi2019} iSF dimers where Davydov splitting is inferred from FC-progression dominated linear spectra and DFT calculations, the Davydov splitting in the structurally well-defined rigid dimer investigated here is \vtnew{directly} confirmed by the presence of 2DES cross-peaks. \\

 Interestingly, the strongest features in the 2D spectrum are the two distinct ESA bands along the excitation axis, where both approximately coincide with the position of the two Davydov components. As shown in \vt{Figure~\ref{fig:fig2}b}(left, middle), when an area at these locations is sampled, \patra{green versus yellow} rectangles in \vt{Figure~\ref{fig:fig2}a}, a strong wavelength-dependent dynamics of the ESA signal is seen. The ESA signal shows a fast decay of 0.19 $\pm$ 0.05 ps and 0.17 $\pm$ 0.05 ps with \patra{green versus yellow} excitation, which is approximately wavelength-independent within the error bars. The rise of the ESA signal is, however, $\sim$3x faster when the lower Davydov component is excited (2.12 $\pm$ 0.32 ps versus 0.60 $\pm$ 0.18 ps). The fitting procedure is described in \vt{Section S2} and summarized in \vt{Table S1, S2}. The fits shown in \vt{Figure~\ref{fig:fig2}b} (left, middle) are freely floated and do not show any trends in the residual (top panels in \vt{Figure~\ref{fig:fig2}b}). In comparison, when the fit is constrained to have the same time constants, a clear residual trend is seen in the 650 nm excitation wavelength region. This is shown in \vt{Figure S3b(right)}. \vt{Figure~\ref{fig:fig2}b}(right) shows the \vt{$T$} dynamics in the dashed yellow rectangle with excitation wavelength at the lower Davydov component as before, but with detection wavelength $\sim$710 nm compared to 760 nm earlier. The $\sim$200 fs decay changes to a $\sim$200 fs rise of the signal, followed by a slower rise on the $\sim$2 ps timescale. Previous\cite{Bansal2022highly} low temporal resolution PP experiments identified the picosecond rise of the ESA signal as the signature of $TT_1$ formation. Here, we find that this process is strongly wavelength dependent, along with an intermediate step that occurs on a $\sim$200 fs timescale, which, however, is not wavelength dependent. Note that we have confirmed that all the above 2DES observations remain unchanged even if the experiments were conducted in a non-polar (weakly polar) DCM solvent (\vt{Section S5}) suggesting that the intermediate states during $S_1 \rightarrow TT_1$ do not involve any significant $CT$ character (\textit{vide infra}). 

\subsection{$TT_1$ formation through a coherently coupled intermediate}
 
To further understand the 200 fs rise and decay of the ESA signal at the detection wavelengths of 710 nm and 760 nm respectively, we ignore the above excitation wavelength dependence and globally fit the 2D spectra with a freely floated three-exponential model. The corresponding rate maps are shown in \vt{Figure~\ref{fig:fig2}c}. A dispersive lineshape along the detection wavelength region is seen for the $\tau_{1}$=0.19 $\pm$ 0.02 ps signal which confirms the visual trends seen in \vt{Figure~\ref{fig:fig2}b}. As may be expected from the neglect of wavelength-dependent dynamics in the global rate map analysis, the \patra{$\tau_{2}$=1.80} $\pm$ 0.21 ps rate map shows a clear trend in the 650 nm excitation wavelength region (\vt{Figure S3c}). Curiously, hardly any rise is seen in the positive GSB band, which is indeed consistent with the observation of positive CP$_U$ between the upper and lower Davydov components (\vt{Figure~\ref{fig:fig2}a}). This is so because the GSB signal strength\cite{Jonas2003,Farrow2008Polarized,Kitney2014Two}, $s^{GSB}$, in a strongly electronically coupled system is already twice that of stimulated emission (SE), $s^{SE}$ -- shared electronic correlations (shared ground state) implies that bleaching one transition also bleaches the other. Therefore, \vt{for a system with positive, upper 2D cross-peak,} no rise in the GSB signal during $LE \rightarrow TT_1$ conversion is to be expected. 

\begin{figure*}[!ht]
	\centering
	\includegraphics[width=6 in]{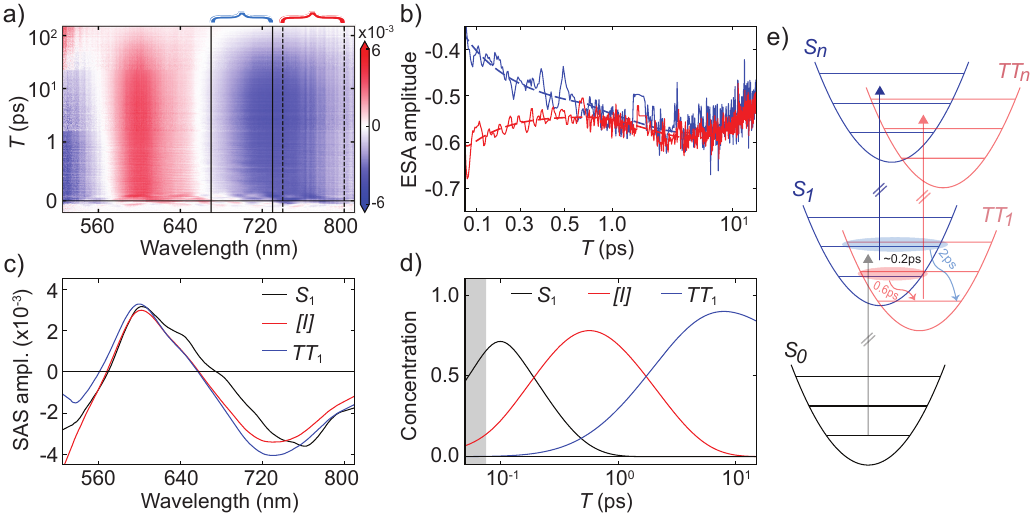}
	\caption{ \footnotesize \textbf{Pump-probe target analysis for the NDI dimer confirms ultrafast formation of the intermediate state}. (\textbf{a}) Spectrally-resolved PP spectrum for the NDI dimer. The marked blue and red regions denote the ESA bands in which $\sim$200 fs rise and decay of the signal is observed along the detection wavelength axis in the 2D rate maps in \vt{Figure~\ref{fig:fig2}c}. (\textbf{b}) Spectrally integrated pump-probe decay traces along $T$ in the blue and red ESA bands. The respective fits from the global analysis of the PP data are shown as dashed lines. (\textbf{c}) Species associated spectra (SAS) obtained after target analysis of the PP data. The corresponding concentration profiles of the associated species are shown in (\textbf{d}) for a sequential target kinetic model: $S_{1}$-$[I]$-$TT_{1}$. The integrated instrument response function (IRF) obtained from the global fit is overlaid as the grey shaded area. The gray band marks the $T$ time point where the integrated IRF is down to 1\%. \vt{Table S1 (rightmost column)} summarizes the pump-probe time constants obtained after the global fits. (\textbf{e}) Summary of all the features in the 2DES rate maps (\vt{Figure~\ref{fig:fig2}c}) and the pump-probe data that are consistent with the target model in panels c,d.
	}
	\label{fig:fig3}
\end{figure*}

The signal-to-noise ratio (SNR) of the 2D signal is $\sim$2.42 due to low pump pulse energies ($\sim$2.1 nJ over 200 nm) and sample extinction coefficient ($\varepsilon$ $\sim$5530 M$^{-1}$cm$^{-1}$). Therefore we confirm the above trends from a separately conducted impulsive PP experiment with better SNR. This also allows us to track the impulsively excited quantum beats, if any, as reporters\cite{Kim2020Non,Heo2025Tracking} for the rich excited state dynamics observed from 2DES measurements. The PP measurements and the corresponding target analysis are presented in \vt{Figure~\ref{fig:fig3}}.

\vt{Figure~\ref{fig:fig3}a} shows the PP spectra as a function of waiting time $T$. The 200 fs rise (fall) of the signal at blue (red) detection wavelengths, marked by dashed versus solid \patra{yellow} squares in \vt{Figure~\ref{fig:fig2}b}, are evident from the blue and red ESA bands in \vt{Figure~\ref{fig:fig3}a}. The corresponding band-integrated $T$ dynamics is shown in \vt{Figure~\ref{fig:fig3}b}, which shows the concomitant rise/fall dynamics. A global fit of the pump-probe data reveals \patra{0.19 $\pm$ 0.02 ps and 1.80 $\pm$ 0.21 ps} as the fastest two time constants, both in good agreement with the 2DES data if the excitation wavelength dependence is ignored. Expectedly, when this global time constant is imposed on the 2D rate maps, a clear trend in the residual is seen for excitation wavelengths corresponding to the lower Davydov component around 650 nm. This is shown in \vt{Figure~S3b,c}. The longest time constant of \patra{73.58 $\pm$ 6.58 ps} is in good agreement with the 70 ps time constant reported earlier\cite{Bansal2022highly} and assigned to triplet-triplet geminate recombination. The details of the global fit are summarized in \vt{Section S1.2.2}. \vt{Figure~\ref{fig:fig3}c,d} show that a minimal target model which allows for the initially excited singlet $S_{1}$ to be converted to an intermediate which then relaxes to the $TT_{1}$ state satisfactorily fits the PP data. The concentration profile in \vt{Figure~\ref{fig:fig3}d} shows that the population of the intermediate species is already maximized by 480 fs. The corresponding species associated spectra (SAS) in \vt{Figure~\ref{fig:fig3}c} show only minimal spectral changes between the initially excited $S_{1}$ and the $TT_{1}$ photoproduct species with strongly overlapping ESA bands. Note that in the earlier pump-probe study\cite{Bansal2022highly} the $TT_{1}$ species was confirmed through a Ruthenium-based triplet sensitizer that transferred triplets to the NDI dimer whose triplet spectrum was measured through a microsecond pump-probe experiment and found to be quite similar to that of $TT_{1}$. The similarity of the spectrum of the intermediate species to the initial $S_{1}$ and the final $TT_{1}$ photoproduct has been reported\cite{Lochbrunner2009,Mauck2016Singlet} previously in case of intermolecular SF. \vt{Combined with our other observations, this suggests that the coherent $[S_1+TT_1]$ intermediate, \vt{denoted as $[I]$ in the Figure~\ref{fig:fig3}}, forms with minimal electronic reorientation during the nuclear evolution away from the initially excited FC geometry}. This assertion is also consistent with the fast formation time of the intermediate species, its excitation wavelength \vtnew{and solvent polarity} independent formation rate (\vt{Figure~\ref{fig:fig2}b} and \vtnew{Section S5, respectively}), and minimal electronic reorientation during this process (\textit{vide infra}). We will come back to this latter point during the discussion of electronic polarization anisotropy. \vt{The overall picture of coherent intermediate formation emerging from the above discussion is summarized in Figure~\ref{fig:fig3}e.}

\subsection{Tracking electronic reorientation during $S_{1} \rightarrow [S_1+TT_1] \rightarrow TT_{1}$}

Two strongly coupled and isoenergetic electronic sites form perfectly mixed excitons and may not show prominent spectral changes during $T$ in case of large spectral overlaps. In such situations, electronic polarization anisotropy can be a quite sensitive probe of electronic dynamics. Presence of positive upper CPs, corresponding well-defined ESA features along the excitation axis in the 2D spectra, fast wavelength-independent formation of an intermediate species (\vt{Figures~\ref{fig:fig2} and \ref{fig:fig3}}), but its strongly wavelength-dependent relaxation to $TT_{1}$ with minimal spectral changes, prompted us to use polarization anisotropy to directly track the transition dipole reorientation that accompanies these steps. 

\begin{figure*}[!ht]
	\centering
	\includegraphics[width=4.5 in]{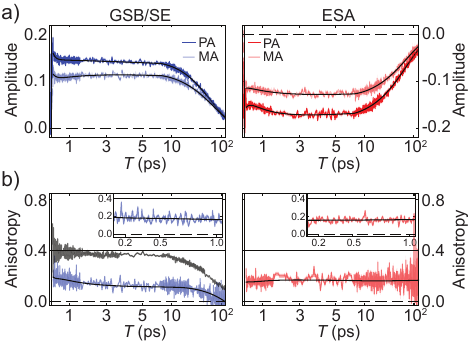}
	\caption{ \footnotesize \textbf{Electronic reorientation during $S_{1}$-$[S_1+TT_1]$-$TT_{1}$ internal conversion}. (\textbf{a}) Parallel (PA) and magic angle (MA) pump-probe transients in a 10 nm GSB/SE band centered at 600 nm and ESA band centered at 725 nm. Fits from the global analysis of MA data in Figure~\ref{fig:fig3} are overlaid starting from 100 fs. The PA fit was obtained by freely floating the same three-exponential fit function. (\textbf{b}) \patra{Polarization anisotropy reconstructed using the respective PA and MA data of the GSB/SE (left) and ESA (right) bands. The anisotropy derived from PA and MA transient fits are also overlaid. The anisotropy of oxazine 720 laser dye is overlaid as a reference (gray trace) in the left figure. Inset shows the zoomed in anisotropy until 1 ps. Horizontal lines in GSB/SE anisotropy mark the average anisotropy of 0.215$\pm$0.008 observed over the $T$ range of 0.12-0.15 ps. The dashed horizontal line in ESA anisotropy at 0.145$\pm$0.003 marks the average ESA anistropy observed over the same $T$ range.}
	}
	\label{fig:fig4}
\end{figure*}
\FloatBarrier
\vt{Figure~\ref{fig:fig4}a,b} shows the parallel (PA) and magic angle (MA) pump-probe transients in the GSB/SE and ESA bands, centered at 600 nm and 725 nm in \vt{Figure~\ref{fig:fig3}a}, respectively. The fit function for the MA transient is the same as that obtained from the global fit in \vt{Figure~\ref{fig:fig3}}. The three-exponential global fit function for the PA case is freely floated to account for polarization-dependent (anisotropic) electronic relaxation. The resulting fits are overlaid on the respective data in \vt{Figure~\ref{fig:fig4}a} and the derived anisotropy is overlaid in bottom panels. Anisotropy of a laser dye, oxazine 720, is also overlaid as a reference for the ideal anisotropy of 0.4 and 0 at early and late pump-probe delays, respectively, expected from an ensemble of isolated electronic transition dipoles in solution. The anisotropy of the reference oxazine sample, averaged over 400-600 fs window, is \patra{$0.396 \pm 0.018$} and \patra{$0.053 \pm 0.012$} in the 500-600 ps window (see also \vt{Figure S9}). In comparison, as shown in \patra{Figure~\ref{fig:fig4}b} (left), the iSF dimer shows a GSB/SE band anisotropy of $0.215 \pm 0.008$ at 120-150 fs, $0.180 \pm 0.002$ at 550-600 fs and $0.144 \pm 0.003$ at 5.5-6 ps (\patra{Table S3}). The change in the GSB/SE band anisotropy to $0.180 \pm 0.002$ in the 600 fs interval likely reflects the SE signal evolution on the excited state as the initial excitation approaches the nuclear geometry of the intermediate while the reduction in anisotropy to $0.144 \pm 0.003$ fs at 6 ps reflects the true GSB anisotropy once the excited state internal conversion to $TT_{1}$ quenches the SE signal.  \\


\textit{A priori}, during internal conversion between species with orthogonally polarized transition dipoles, such as $S_{1} \rightarrow TT_{1}$ conversion in acenes\cite{Michl2010} with transition dipole $\hat{\mu}_{S_0 \rightarrow S_1}$ along the short axis and $\hat{\mu}_{TT_1 \rightarrow TT_n}$ polarized along the long axis, anisotropy\cite{Jonas1996Pump} $r(\theta_{ij}) = (3 \cos^2(\theta_{ij}) - 1)/{5}$, where $i, j$ are the initial and final species and $\theta$ is the relative angle between their transition dipoles, can be expected to change from 0.4 to -0.2. A similar anisotropy of approximately -0.2 has been recently reported\cite{Vauthey2025} by Vauthey and co-workers in the context of a $CT$ donor-acceptor system. In the current context, species $i$ is the initially excited singlet while $j$ could be the intermediate or later the relaxed $TT_{1}$ species. The above changes in anisotropy can be directly related to $\Delta\theta_{ij}$ change of $3.53^\circ$ and $3.49^\circ$ at 600 fs and 6 ps, respectively, in the electronic transition dipole orientation relative to 200 fs, suggesting minimal electronic reorientation on the excited state during iSF. However, a minimal change in the anisotropy in the GSB/SE band may not reflect the true electronic reorientation because of weak SE signal strength. This is so because ultrafast quenching of the SE signal due to iSF implies that the total anisotropy $r_{tot}$ in the GSB/SE band, given by $r_{tot} = (s^{GSB} \cdot r_{GSB} + s^{SE} \cdot r_{SE})/({s^{GSB} + s^{SE}})$, will be sensitive to the GSB signal only. In contrast, anisotropy in the ESA band, shown in \patra{Figure~\ref{fig:fig4}b} (right), reflects excited state changes and provides the opportunity to exclusively track electronic reorientation during $S_1 \rightarrow TT_1$ conversion without any contamination from the ground state anisotropy. Surprisingly, minimal changes in the anisotropy are observed in the ESA band as well. The expected anisotropy change during $S_1 \rightarrow TT_1$ conversion, that is parallel to orthogonally polarized initial and final states transitions, $\mu_{S_0 \rightarrow S_1} \parallel \mu_{S_1\rightarrow S_n}$ and $\mu_{S_0 \rightarrow S_1} \perp \mu_{TT_1\rightarrow TT_n}$, respectively, is 0.4 to -0.2. Compared to this, the average $r_{ESA}$ in the three time windows is $0.145 \pm 0.013$, $0.150 \pm 0.014$ and $0.152 \pm 0.016$, respectively (\patra{Table S3}), all within the error bars of each other. Note that we have confirmed that the anisotropy in the ESA band stays $\sim$0.15 with $T$ across the band and not sensitive to the position of the chosen band. This observation directly reports negligible electronic reorientation during the nuclear evolution towards the intermediate $[S_1+TT_1]$ and its subsequent relaxation to the $TT_{1}$ state. It also raises the question that what causes the observed ESA anisotropy to deviate so much from the expected limits of the anisotropy during iSF ? A similar question also arises for the GSB/SE band anisotropy of 0.215 at 200 fs which lies significantly below the expected 0.4 for an isolated transition dipole. \\

To better understand the limits of polarization anisotropy observed in the experiments, results\cite{Jonas2003} from purely excitonic dimers with orthogonal transition dipoles provide a useful starting point. Although the case of iSF dimer is more complicated due to $CT$ and $TT_1$ states which may borrow oscillator strength from the $LE$ states, in the impulsive limit where all states are excited, $r_{GSB}$ of both dimers will be the same because oscillator strength arises from the $LE$ states in both cases. Strongly correlated electronic states with \textit{orthogonal} transition dipoles which share a common ground state, such that the signal strengths are $s^{GSB} = 2 s^{SE}$, show\cite{Jonas2003, Jonas2011} $r_{GSB}$ between 0.1 in the impulsive excitation limit to 0.4 in the single transition dipole excitation limit. In comparison, the SE signal, assuming no electronic coherence signal pathways, shows $r_{SE}$ ($T=0$) of 0.4 corresponding to SE from a transition dipole that is maximally aligned with the pump polarization. \patra{Section S4} extends these results to a generalized excitonic dimer with any angle $\theta_{AB}$ between the transition dipoles associated with sites $A$ and $B$. \\

In case of orthogonal dipoles, the GSB anisotropy monotonically increases\cite{Jonas2011} from 0.1 under impulsive excitation to 0.4 under (non-impulsive) single transition dipole excitation. However, for the generalized dimer model, \patra{Figure~S5} shows that the monotonic increase\cite{Jonas2011} in $r_{GSB}$ from 0.1 to 0.4 is \textit{reversed} at intermediate values of $\theta_{AB}$. This is due to the interference between the transition dipoles which is absent for the orthogonal case. For $\theta_{AB}$ = 24$^o$, the expected GSB anisotropy with impulsive excitation that covers all states is \patra{$r_{GSB}$ = 0.35} compared to the observed value of 0.215 at 200 fs. From \patra{Figure~S5} the reduced anisotropy can be understood as due to the partial laser coverage of the upper Davydov component at $\sim$600 nm as well as to its loss of oscillator strength that gets distributed to higher lying $CT$ states due to large $LE-CT$ mixing. As we noted earlier, indeed the TD-DFT calculations$^{36}$ from Zhu and co-workers on a contorted PDI dimer show that the singlet oscillator strength is significantly re-distributed among higher-lying $CT$ and singlet configurations (see Section S8 of ref.~\cite{Conrad2019Controlling}). The observation of reduced GSB anisotropy could also mean that  the relative orientation of the NDI chromophores in the dimer changes significantly between the crystal and the solution. This possibility seems unlikely for a rigid dimer. Note that even if the SE signal was considered in the estimation of GSB/SE band anisotropy, the total anisotropy will only increase to deviate more from the observed anisotropy. \patra{Figure S7} shows that the amount of ESA contribution in the GSB/SE band is only $\sim10\%$ and not sufficient to explain the reduced ansisotropy in the GSB/SE band. Overall, reduced anisotropy in the GSB/SE band complements the evidence of strongly coupled iSF dimer with shared ground state correlations, consistent with the positive, upper cross peak reported by the 2DES spectra (\vt{Figure~\ref{fig:fig2}}).\\

As we noted earlier, the ESA band provides the opportunity to look at purely excited state dynamics. The expected range of $r_{ESA}$ from 0.4 to -0.2, for 0\% to 100\% $\hat{\mu}_{TT_1 \rightarrow TT_n}$ character (100\% to 0\% $\hat{\mu}_{S_1\rightarrow S_n}$ character) of the ESA band, respectively, suggests that the experimentally observed $r_{{ESA}}$ of $\sim0.15$ represents an interesting case where the ESA band is polarized in between $\hat{\mu}_{S_1 \rightarrow S_n}$ and  $\hat{\mu}_{TT_1 \rightarrow TT_n}$ transitions. This is not surprising given the nearly identical spectra of the singlet and the triplet species (Figure\ref{fig:fig3}). Assuming both transitions are present within the ESA band with strengths $\kappa$ and $\gamma$, respectively, the total anisotropy in the ESA band can be expressed\cite{Qian2003Role} as $r_{ESA} = (\kappa^2 r_{S_1 \rightarrow S_n} + \gamma^2 r_{TT_1 \rightarrow TT_n})/({\kappa^2 + \gamma^2 })$. Equating this to the observed anisotropy predicts that the ESA band consists of $\sim$60-40 ratio of $\hat{\mu}_{S_1 \rightarrow S_n}$ and  $\hat{\mu}_{TT_1 \rightarrow TT_n}$ polarized transitions. The intermediate polarization of the ESA band with no significant change with $T$ directly reports that the $S_1 \rightarrow [S_1+TT_1] \rightarrow TT_1$ internal conversion process involves surprisingly minimal electronic reorientation, maintains the strong electronic correlations between the chromophore sites with significant singlet-triplet mixing throughout $TT^1$ formation and relaxation. \vtnew{To best of our knowledge, the surprising lack of electronic reorientation during iSF has not been reported before and suggests that synthetic design should aim for minimizing such mixing to prolong the lifetime of high-spin correlated triplets. Such mixing also needs to be accounted for in the electronic structure methods that aim to predict the high-spin state dynamics that follows $(TT_1)$ generation.} \\

At this point, the question arises as to what is the nature of the intermediate $[S_1+TT_1]$. Measurements of weakly emissive sharp emission profiles in the excitation–emission spectra (\vt{Figure~\ref{fig:fig1}b,c}) suggest intensity borrowing arising from $LE-CT$ mixing is likely. However, the position of these weakly emissive states is at least 2043~cm$^{-1}$ (75~nm) above the lower Davydov component at 15625~cm$^{-1}$ (640~nm). Signatures of NDI radical cation and anion bands\cite{Bansal2022highly,Ajayakumar2012Core} in the PP spectra are absent. \patra{Figure S8} overlays the cation and anion bands with the NDI dimer PP spectrum where no such bands are obvious. Thus no significant $CT$ character may be expected in the intermediate state. This assertion is further supported by our observations of solvent polarity independent rates (\vt{Figure~S6}) for both steps -- the intermediate formation as well as its wavelength dependent relaxation to $TT_1$. SF through intermediates with minimal to absent $CT$ character have been reported\cite{Lochbrunner2009,Guldi2019} previously. To further investigate this question, we measured impulsive vibrational quantum beat spectra during the $S_1$--$[S_1+TT_1]$--$TT_1$ electronic relaxation. 

\begin{figure*}[!ht]
	\centering
	\includegraphics[width=6 in]{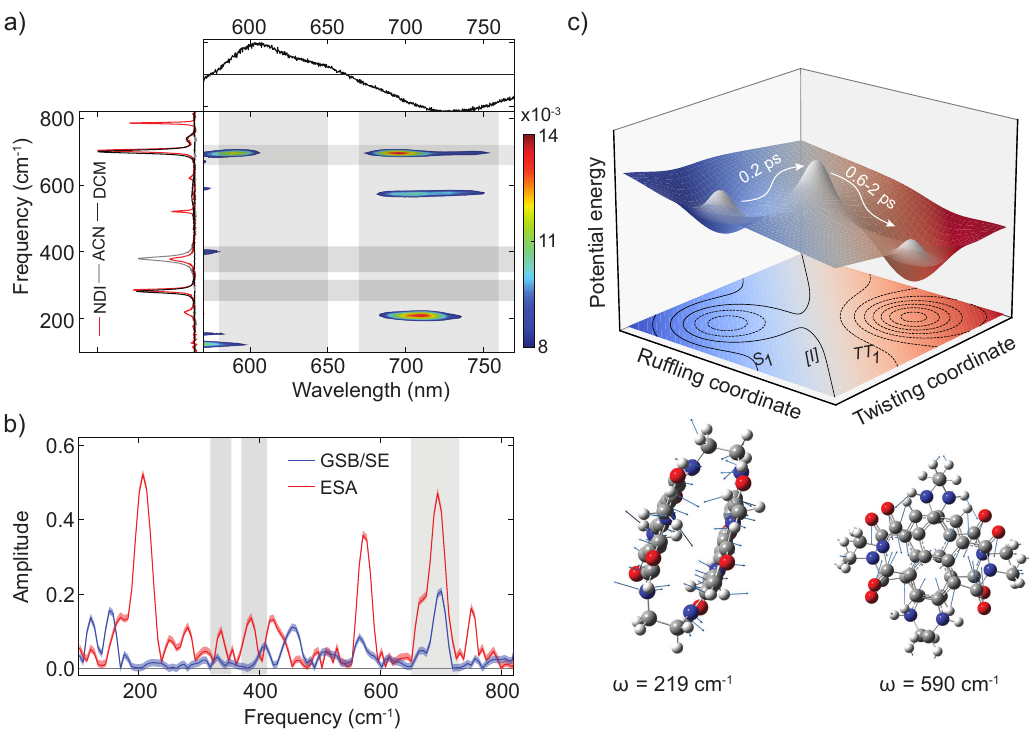}
	\caption{\footnotesize \textbf{Enhanced quantum beats in the \patra{$TT_1$} photoproduct correspond to inter-chromophore nuclear motions.} (\textbf{a}) Quantum beat spectrum as a function of detection wavelength. The top panel shows the PP spectrum at $T$ = 1 ps, while the side panel shows the ground state Raman spectrum of the NDI dimer excited at 785 nm along with the non-resonant Raman spectra of the two solvents ACN and DCM. The lowest contour is at 20\%. Quantum beats in the PP data arising from the  non-resonant Raman scattering from the solvent are marked as the horizontal grey band. (\textbf{b}) The vertical gray bands in panel a are spectrally integrated and overlaid for the GSB/SE (blue) and ESA (red) bands. Vertical gray bands mark the solvent bands, same as horizontal bands in panel a. (\textbf{c}) Schematic for the excimer-like intermediate formation due to large excited state nuclear displacements along the theoretical \patra{219} cm$^{-1}$ and \patra{590} cm$^{-1}$ modes that correspond to inter-chromophore ruffling and twisting motions, respectively (Section S1,S3).
	}
	\label{fig:fig5}
\end{figure*}
\FloatBarrier
\vt{Figure~\ref{fig:fig5}a} shows the quantum beat spectrum as a function of detection wavelength obtained from the PP data in \vt{Figure~\ref{fig:fig3}}. The PP spectra shows two modes at 210~cm$^{-1}$ and 575~cm$^{-1}$ that are prominently seen on the $TT_1$ photoproduct and nearly absent in the ground electronic state. This can also be seen in \vt{Figure~\ref{fig:fig5}b} which overlays the spectrally integrated GSB and ESA bands which are denoted by the gray vertical bands in \vt{Figure~\ref{fig:fig5}}a. The ground state Raman spectrum for the NDI dimer, overlaid as left vertical panel, shows only weak Raman peaks corresponding to the quantum beats at $\sim$210~cm$^{-1}$ and $\sim$575~cm$^{-1}$ and consistent with their near absence in the GSB band observed in the PP experiments. Further, \patra{Figure~S4a} shows a similar plot as \vt{Figure~\ref{fig:fig5}a} for the NDI monomer. Exactly similar trends are seen in the NDI monomer as well, that is, very weak to absent modes at $\sim$180~cm$^{-1}$ and $\sim$580~cm$^{-1}$ on the ground electronic state. The absence of these prominent modes on the ground electronic state of the NDI monomer is also consistent with the Raman spectrum of related NDI derivatives reported\cite{SosorevPCCP2023} previously (see Figure 5 of ref. \cite{SosorevPCCP2023}). \patra{Figure~S4c} also compares the mode frequencies between the NDI monomer and the dimer where shifts of 30~cm$^{-1}$ for the 210~cm$^{-1}$ mode and 5~cm$^{-1}$ for the 575~cm$^{-1}$ are seen. These shifts likely arise due to the constrained skeleton of the dimer. \\

Normal mode analysis of the two most prominent modes seen on the $TT_1$ photoproduct suggest that multiple modes around $\sim$200-300 cm$^{-1}$ correspond to inter-chromophore ruffling motions while those around $\sim$520-600 cm$^{-1}$ correspond to inter-chromophore twisting motions (see \patra{Table S4}). Both motions are expected to affect the intermolecular orbital overlaps.  Joo and co-workers have shown\cite{Kim2020Non,Heo2025Tracking} that enhanced vibrational quantum beats on the excited state can directly report on the nuclear displacements that are the most displaced during internal conversion. Prominent inter-chromophore ruffling and twisting motions, observed as enhanced $\sim$210~cm$^{-1}$ and $\sim$575~cm$^{-1}$ quantum beats on the $TT_1$ photoproduct, suggests significant nuclear displacements along these coordinates during internal conversion, likely resulting in an excimer-like $[S_1+TT_1]$ intermediate which then relaxes to the $TT_1$ state. Excimer formation\cite{Lim2000} in \patra{naphthalenediimide} dimers is indeed sensitive to relative orientation and expected to be facile in case of eclipsed dimers. A functional role for modulation of inter-chromophore orbital overlaps was recently implicated\cite{Vener2021Non} in charge transport in NDI crystals. Low-frequency inter-chromophore motions and their effect on electron-phonon couplings are also known in pentacene polycrystalline films where Peierls couplings have been implicated\cite{Bhattacharyya2023Low,Tempelaar2018Vibronic,Huo2017} in assisting SF in multiple reports. \vtnew{Notably, similar low frequency ($\sim$ 200 cm$^{-1}$) intermolecular out-of-plane vibrations have also been implicated\cite{Kim2021, Kim2022} in excimer intermediate formation in perylenebisimide (PBI) dimers. Similarly, $\sim$500cm$^{-1}$) ring breathing and deformation motions have been implicated\cite{KimWurthner2022} in the iSF process in PBI trimers. Our observations of enhanced excited state quantum beats along the inter-chromophore ruffling and twisting coordinates suggest that such structural dynamics may be more general across various iSF platforms.} \\

Large admixtures of $LE-CT$ states created by orbital overlaps can lead\cite{SpanoExcimer2022} to excimer-like states. To further probe the $CT$ character in the intermediate, we performed 2DES experiments and rate map analysis on the NDI dimer dissolved in a non-polar (weakly polar) DCM solvent. The resulting spectral features and observations are same as in the polar solvent mixture (highly polar ACN with DCM). The corresponding 2DES spectrum at $T=$ \patra{0.2 ps} and its excitation wavelength dependence are shown in \vt{Figure S6}, and suggest that the $[S_1+TT_1]$ intermediate is likely to have only a weak $CT$ character not sufficient to significantly alter the wavelength dependent $TT_1$ formation kinetics. Excimer intermediate in SF with weak to absent $CT$ character has been previously reported\cite{Guldi2019,Lochbrunner2009}. However, excimer-mediated SF has also been reported in systems where the $CT$ character in the excimer intermediate is tuned through molecular packing\cite{Mauck2016Singlet} and \vtnew{solvent polarity\cite{KimWurthner2022} and larger $CT$ character in the excimer intermediate is correlated with faster $TT_1$ formation. Solvent polarity independent rates suggest only weak $CT$ character is involved and therefore we assign the intermediate to an excimer-like species where we confirm the excimer-like nature of the intermediate directly by analyzing the nuclear motions corresponding to the quantum beats that are enhanced in the $TT_1$ photoproduct, an approach suggested\cite{Kim2020Non,Heo2025Tracking} by Joo and co-workers. }\\

The overall picture of iSF in NDI dimers that emerges from our observations is summarized in \vt{Figure~\ref{fig:fig5}c}. The initial $\sim$200 fs rise of the intermediate followed by a slower rise of the ESA signal is reminiscent of the vibronically coherent [$S_1$ + $TT_1$] intermediate formation reported in tetracene thin films\cite{Stern2017Vibronically} by Friend and co-workers. Similar coherently coupled $LE-CT$ intermediates have been recently implicated\cite{Lin2022Accelerating,Hong2022Ultrafast} in symmetry-breaking charge transfer dyads. Interestingly, only minor changes in electronic anisotropy are seen during the evolution of the $S_1$ state suggesting minimal electronic reorientation during the nuclear evolution that leads to the intermediate formation. The accompanying nuclear motions do not seem to localize the initially prepared state and maintain the strong electronic correlations between the sites (shared ground state). A complete localization of the initial photoexcitation would have resulted in GSB anisotropy increasing to $\sim$0.4 during intermediate formation. This suggests a picture where the [$S_1$+$TT_1$] intermediate is coherently coupled to the $LE$ and $TT_1$ states with strong correlations between the sites, explaining its wavelength independent and fast formation -- 30\% of the [$S_1$+$TT_1$] intermediate is already formed within the IRF window in \vt{Figure~\ref{fig:fig3}d}. Observations of upper 2D cross peak and reduced anisotropy in the GSB/SE band are also both consistent with the above picture. Notably, $LE-TT$ mixing with borrowed oscillator strength in the $TT_1$ state resulting in direct photoexcitation has been reported\cite{Musser2024, Turner2017,Zhu2017,Zhu2012} in several acene thin films. Our observations show that similar effects may indeed be present in strongly coupled rigid iSF dimers even in the absence of thin films. \vtnew{Tracking electronic dynamics accompanying structural evolution directly shows that the $TT^1$ photoproduct maintains significant singlet character which likely accelerates its fast recombination. }




\section*{CONCLUSIONS}

The chemical space for iSF chromophores is rapidly expanding while detailed mechanistic insights into the electronic and nuclear motions involved during the correlated triplet pair \patra{formation} are rare. NDI-based cyclophanes have been recently introduced\cite{Bansal2022highly} as a new class of iSF systems where a combination of contortion and rotation makes iSF a favourable pathway. Through a combination of polarization-controlled  2DES and pump-probe spectroscopy, we have presented a comprehensive mechanistic investigation of the correlated triplet pair formation in these dimers with conclusive evidence for an excimer-like intermediate state whose $\sim$200 fs formation is excitation wavelength independent but subsequent $\sim$0.6-2 ps relaxation to $TT_1$ is strongly dependent on which Davydov component of the dimer is excited -- a $\sim$3x faster triplet formation rate is observed when exciting the lower Davydov component. Enhanced vibrational quantum beats in the $TT_{1}$ photoproduct implicate inter-chromophore nuclear motions -- twisting and ruffling -- as key drivers of the intermediate formation and its subsequent relaxation to $TT_1$. Shared electronic correlations between the chromophore sites and surprisingly minimal electronic reorientation during these steps directly reported by \patra{2DES} cross-peaks and polarization anisotropy suggest a coherent picture of iSF where strong electronic correlations between the chromophores \vtnew{and singlet-triplet mixing} is maintained during the $S_1$ – [$S_1$+$TT_{1}$] – $TT_{1}$ electronic evolution. The surprising lack of electronic reorientation during iSF in strongly coupled dimers has not been reported before and motivates synthetic design and theoretical models which balance strong correlations between the triplets while diminishing singlet-triplet mixing with nuclear evolution. Predictions of high-spin state dynamics arising from correlated triplets also need to account for singlet-triplet mixing during \patra{$TT_1$} relaxation. \vtnew{In the context of tracking\cite{KimWurthner2022} structural dynamics during iSF, our observations demonstrate that tracking electronic dynamics provides complementary synthetic and theoretical insights for advancing iSF systems. }Our study also introduces impulsive electronic polarization anisotropy\cite{Hauer2025} as a powerful tool that can directly track electronic reorientation dynamics during singlet fission lending powerful mechanistic insights necessary for refining the current theoretical models for singlet fission. \\



\section*{ASSOCIATED CONTENT}
\textbf{Supporting Information} \\
Experimental details, including synthesis and geometry optimizations, optical layout, spectroscopic measurement procedures, as well as wavelength-dependence analysis. \\

\textbf{Notes}
The authors declare no competing interests.

\section*{ACKNOWLEDGEMENTS}
{\footnotesize V.T. thanks Prof. Jyotishman Dasgupta, Department of Chemical Sciences, TIFR Mumbai for initiating this collaboration and for the numerous fruitful discussions. A.B. acknowledges the Prime Ministers' Research Fellowship, MoE, India. S.P. acknowledges the MHRD research fellowship from the Indian Institute of Science. P.M. acknowledges SERB grant no. CRG/2021/008494 for partial funding. P. M. thanks Dr. Deepak Bansal for his valuable initial inputs in the synthesis of the NDI dimers. D.G. acknowledges the support of the SERB-ANRF Fellowship (CRG/2023/001806) and the computational facilities provided by IACS. S.S. acknowledges funding support from the DST-INSPIRE program through a Junior Research Fellowship. V.T. acknowledges funding from SERB-ANRF under grant sanction number CRG/2023/000327, from ISRO-STC under grant sanction number ISTC/CSS/VT/505, and from the Quantum Research Park (QuRP) funded by the Government of Karnataka.}

\renewcommand{\refname}{REFERENCES}   
\renewcommand{\bibname}{REFERENCES}   

\bibliography{NDIRefs}

\end{document}